\documentclass[letterpaper, 10 pt, conference]{ieeeconf}  
\IEEEoverridecommandlockouts
\usepackage{balance}
\usepackage{color}
\usepackage{caption}
\usepackage{subcaption}
\usepackage{enumerate}
\usepackage{cite}
\usepackage{empheq}
\usepackage{mathrsfs}

\usepackage{mathtools}

\usepackage{tikz}
\usepackage{circuitikz}


\makeatletter
\pgfcircdeclarebipole{}{\ctikzvalof{bipoles/interr/height 2}}{spst}{\ctikzvalof{bipoles/interr/height}}{\ctikzvalof{bipoles/interr/width}}{

    \pgfsetlinewidth{\pgfkeysvalueof{/tikz/circuitikz/bipoles/thickness}\pgfstartlinewidth}

    \pgfpathmoveto{\pgfpoint{\pgf@circ@res@left}{0pt}}
    \pgfpathlineto{\pgfpoint{.6\pgf@circ@res@right}{\pgf@circ@res@up}}
    \pgfusepath{draw}   
}

\def\pgf@circ@spst@path#1{\pgf@circ@bipole@path{spst}{#1}}
\tikzset{switch/.style = {\circuitikzbasekey, /tikz/to path=\pgf@circ@spst@path, l=#1}}
\tikzset{spst/.style = {switch = #1}}
\makeatother

\makeatletter
\let\proof\@undefined                        
\let\endproof\@undefined                  
\makeatother
\usepackage{graphicx,amssymb,amstext,amsmath,amsthm}

\usepackage[bookmarks=true]{hyperref}
\usepackage{algorithm,algorithmicx,algpseudocode}
\algnewcommand{\algorithmicgoto}{\textbf{go to}}%
\algnewcommand{\Goto}[1]{\algorithmicgoto~\ref{#1}}%
\algnewcommand{\LineComment}[1]{\Statex \(\triangleright\) #1}
\algnewcommand{\LineCommentN}[1]{\Statex \hspace{1cm}\(\triangleright\) #1}

\usepackage{multirow}
\usepackage{stfloats}

\newtheorem{prop}{Proposition} 

\newtheorem{thm}{Theorem}
	\newtheorem{assumption}{Assumption}
\newtheorem{lem}{Lemma}
\newtheorem{defn}{Definition}
\newtheorem{rem}{Remark}
\newtheorem{problem}{Problem}

\usepackage{setspace}

\let\oldbibliography\thebibliography
\renewcommand{\thebibliography}[1]{%
  \oldbibliography{#1}%
}

\newcommand{\yong}[1]{{\color{black} #1}}
\newcommand{\moh}[1]{{\color{black} #1}}
\newcommand{\yongn}[1]{{\color{black} #1}}

\begin{document}

\title{\LARGE \bf Interval Observers for Simultaneous State and Model Estimation\\ of Partially Known Nonlinear Systems} 

\author{%
Mohammad Khajenejad, Zeyuan Jin, Sze Zheng Yong\\
\thanks{
M. Khajenejad and S.Z. Yong are with the School for Engineering of Matter, Transport and Energy, Arizona State University, Tempe, AZ, USA (e-mail: \{mkhajene, szyong\}@asu.edu).}
\thanks{This work is partially supported by NSF grant CNS-1932066.}
}

\maketitle
\thispagestyle{empty}
\pagestyle{empty}

\begin{abstract}
 We study the problem of designing interval-valued observers that simultaneously estimate the system state and learn an unknown dynamic model  
for partially unknown nonlinear systems with dynamic unknown inputs and bounded noise signals. Leveraging affine abstraction methods and the existence of nonlinear decomposition functions, as well as applying our previously developed data-driven function over-approximation/abstraction approach to over-estimate the unknown dynamic model, 
our proposed observer recursively computes the maximal and minimal elements of the estimate intervals that are proven to contain the true augmented states. Then, using observed output/measurement signals, the observer iteratively shrinks the intervals by eliminating estimates that are not compatible with the measurements. Finally, given new interval estimates, the observer updates the over-approximation of the unknown model dynamics. 
Moreover, we provide sufficient conditions for uniform boundedness of the sequence of estimate interval widths, i.e., stability of the designed observer, in the form of tractable  (mixed-)integer programs with finitely countable feasible sets.  
 \end{abstract}
\section{Introduction}
\emph{Motivation}. Motivated by the need to ensure safe and smooth operation in many safety-critical engineering applications such as fault detection, urban transportation, attack (unknown input) mitigation and detection in cyber-physical systems and aircraft tracking \cite{liu2011robust,yong2016tcps,yong2018simultaneous}, 
robust algorithms for state and input estimation have been recently applied to derive compatible estimates of states and unknown inputs. Particularly, set/interval membership approaches have been broadly used to guarantee hard accuracy bounds in safety-critical bounded-error settings. Further, in practical systems, the existence of potentially dynamic unknown inputs with unknown dynamics makes the entire setting a partially unknown system. Thus, the development of appropriate data-driven methods that can deal with the noisy estimated data obtained form set/interval membership approaches to estimate/approximate/abstract unknown system models is a critical and interesting problem.   

\emph{Literature review}.
 Multiple approaches have been proposed in the literature to design set/interval observers  \cite{jaulin2002nonlinear,kieffer2004guaranteed,moisan2007near,bernard2004closed,raissi2010interval,raissi2011interval,mazenc2011interval,mazenc2013robust,wang2015interval,efimov2013interval,zheng2016design,mazenc2014interval,ellero2019unknown,yong2018simultaneous,khajenejad2019simultaneous,khajenejadasimultaneous,khajenejad2020nonlinear}, including linear time-invariant (LTI) \cite{mazenc2011interval},  linear parameter-varying (LPV) \cite{wang2015interval,ellero2019unknown}, Metzler and/or  partial linearizable \cite{raissi2011interval,mazenc2013robust}, cooperative \cite{raissi2010interval,raissi2011interval}, Lipschitz nonlinear \cite{efimov2013interval}, monotone nonlinear \cite{moisan2007near,bernard2004closed} and uncertain nonlinear \cite{zheng2016design} systems. 
However, the aforementioned works either do not consider unknown inputs ({i.e., input, disturbance, attack or noise signals with unknown dynamics to be reconstructed/estimated}) {\cite{jaulin2002nonlinear,kieffer2004guaranteed,moisan2007near,bernard2004closed,raissi2010interval,raissi2011interval,mazenc2011interval,mazenc2013robust,wang2015interval,efimov2013interval,zheng2016design,mazenc2014interval}, or 
 the (potentially unbounded) unknown inputs do not affect the output (measurement) equation \cite{ellero2019unknown}. Considering systems where both state and output equations are affected by arbitrary unknown inputs, 
the problem of simultaneously designing state and unknown input ``set-valued" observers has been studied in our previous works for LTI \cite{yong2018simultaneous}, LPV \cite{khajenejad2019simultaneous}, switched linear \cite{khajenejadasimultaneous} and nonlinear \cite{khajenejad2020nonlinear} systems with bounded-norm noise, while in our recent work \cite{khajenejad2020simultaneous}, we particularly designed ``interval-valued" observers for Lipschitz mixed-monotone nonlinear systems affected by arbitrary unknown inputs.   
 
 On the other hand, considering set-valued uncertainties, data-driven approaches 
that use sampled/observed input-output data to \emph{abstract} or over-approximate 
unknown dynamics using a bounded-error setting,  
have gained increased popularity over the last few years \cite{Milanese2004SetMI,canale2014nonlinear,zabinsky2003optimal,beliakov2006interpolation,calliess2014conservative}. 
The general objective of such data-driven methods is to find \emph{a set of known systems} that share the most properties of interest with the unknown system dynamics \cite{Milanese2004SetMI,canale2014nonlinear}, under the assumption that the unknown dynamics is univariate Lipschitz continuous \cite{zabinsky2003optimal}, multivariate Lipschitz continuous \cite{beliakov2006interpolation} or H\"{o}lder continuous \cite{calliess2014conservative}.
Nonetheless, to our knowledge, these approaches do not explicitly deal with noise/disturbance and their effect on the abstraction, which is especially critical when dealing with ``estimated" data. Hence, in our previous work \cite{Jin2020datadriven}, we generalized the aforementioned data-driven approaches to develop an abstraction approach that can use the noisy sampled/observed/estimated data to over-approximate the unknown Lipschitz continuous dynamics with upper and lower functions.

 \emph{Contributions.} 
 The goal of this paper is to bridge between model-based set/interval-valued observer design approaches, e.g., in \cite{jaulin2002nonlinear,kieffer2004guaranteed,moisan2007near,bernard2004closed,raissi2010interval,raissi2011interval,mazenc2011interval,mazenc2013robust,wang2015interval,efimov2013interval,zheng2016design,mazenc2014interval,ellero2019unknown,yong2018simultaneous,khajenejad2019simultaneous,khajenejadasimultaneous,khajenejad2020nonlinear} and data-driven function approximation methods, e.g., in \cite{Milanese2004SetMI,canale2014nonlinear,zabinsky2003optimal,beliakov2006interpolation,calliess2014conservative}, to design interval-valued observers for nonlinear dynamical systems with bounded noise and \emph{dynamic unknown inputs}, where the state and observation vector fields belong to a fairly general class of nonlinear functions and the unknown input dynamics is governed by an \emph{unknown input function}.  
 By extending the observer design approach in \cite{khajenejad2020simultaneous}, 
 we include a crucial \emph{update step}, 
 where starting from the intervals from the propagation step, the framers are iteratively updated by computing their intersection with the augmented state intervals that are compatible with the observations, resulting in the decreased width updated framers, which leads to obtain tighter intervals. 
   
Moreover, by assuming a mild assumption of Lipschitz continuity for the unknown input functions and applying our previous data-driven function approximation/abstraction approach \cite{Jin2020datadriven} to recursively over-approximate the unknown input function from the noisy estimated intervals/data obtained from the update step, as well as leveraging the combination of nonlinear decomposition/bounding functions \cite{yang2019sufficient,coogan2015efficient,yang2019tight,khajenejad2020simultaneous} and affine abstractions \cite{singh2018mesh}, we prove that our observer is correct, i.e., the framer property \cite{mazenc2013robust} holds and our estimation/abstraction of the unknown input model becomes more precise/tighter over time. More importantly, we provide 
sufficient conditions, in the form of tractable (mixed-)integer programs with finitely countable feasible sets, for the stability of our observer {(i.e., the uniform boundedness of the sequence of estimate interval widths)}. Further we compute uniformly bounded and convergent upper intervals for the sequence of estimates and derive their steady-state values. 
\section{Preliminaries}

{\emph{{Notation}.}} $\mathbb{R}^n$ denotes the $n$-dimensional Euclidean space and $\mathbb{R}_{+{+}}$ positive real numbers. 
For vectors $v,w \in \mathbb{R}^n$ and a matrix $M \in \mathbb{R}^{p \times q}$, $\|v\|\triangleq \sqrt{v^\top v}$ and $\|M\|$ denote their (induced) $2$-norm, \yongn{and} 
$v \leq w$ is an element-wise inequality. 
Moreover, the transpose, 
Moore-Penrose pseudoinverse{, $(i,j)$-th element} 
and rank of $M$ are given by $M^\top$, 
$M^\dagger${, $M_{i,j}$} 
and ${\rm rk}(M)$, while $M_{(r:s)}$ is a sub-matrix of $M$, consisting of its $r$-th through $s$-th rows. We call $M$ a non-negative matrix, i.e., $M \geq 0$, if $M_{i,j} \geq 0, \forall i \in \{1\dots p\},\forall j \in \{1 \dots q \}$. Also, $M^+,M^{++} \in \mathbb{R}^{p \times q}$ are defined as $M^+_{i,j}=M_{i,j}$ if $M_{i,j} \geq 0$, $M^+_{i,j}=0$ if $M_{i,j} < 0$, $M^{++}=M^+-M$ and $|M| \triangleq M^++M^{++}$. Furthermore, $r=\textstyle{rowsupp}(M) \in \mathbb{R}^p$, where $r_i=0$ if the $i$-th row of $A$ is zero and $r_i=1$ otherwise, $\forall i \in \{1\dots p \}$. For a symmetric matrix $S$, $S \succ 0$ and $S \prec 0$ ($S \succeq 0$ and $S \preceq 0$) are
 positive and negative (semi-)definite, respectively.
Next, we introduce some definitions and related results that will be useful throughout the paper.

\begin{defn}[Interval, Maximal and Minimal Elements, Interval Width]\label{defn:interval}
{An (multi-dimensional) interval {$\mathcal{I}  \subset 
\mathbb{R}^n$} is the set of all real vectors $x \in \mathbb{R}^n$ that satisfies $\underline{s} \le x \le \overline{s}$, where $\underline{s}$, $\overline{s}$ and $\|\overline{s}-\underline{s}\|$ are called minimal vector, maximal vector and width of $\mathcal{I}$, respectively}.
\end{defn}
\begin{prop}\cite[Lemma 1]{efimov2013interval}\label{prop:bounding}
Let $A \in \mathbb{R}^{m \times n}$ and $\underline{x} \leq x \leq \overline{x} \in \mathbb{R}^n$. Then\moh{,} $A^+\underline{x}-A^{++}\overline{x} \leq Ax \leq A^+\overline{x}-A^{++}\underline{x}$. As a corollary, if $A$ is non-negative, then $A\underline{x} \leq Ax \leq A\overline{x}$.
\end{prop}
\begin{defn}[Lipschitz Continuity]\label{defn:lip}
A vector field $q(\cdot):\mathbb{R}^n \rightarrow \mathbb{R}^m$ is $L_q$-Lipschitz continuous on $\mathbb{R}^n$, if  $\exists L_q \in \mathbb{R}_{+{+}}$, such that $\|q(\zeta_1)-q(\zeta_2)\| \leq L_q \|\zeta_1-\zeta_2\|$, $ \forall \zeta_1,\zeta_2 \in \mathbb{R}^n$.
\end{defn}
\begin{defn}[Mixed-Monotone Mappings and Decomposition Functions]\cite[Definition 4]{yang2019sufficient}\label{defn:mixed-monotone}
A mapping $f:\mathcal{X} \subseteq \mathbb{R}^n \rightarrow \mathcal{T} \subseteq \mathbb{R}^m$ is mixed monotone if there exists a decomposition function $f_d:\mathcal{X} \times \mathcal{X} \rightarrow \mathcal{T}$ satisfying: 
\begin{enumerate}
\item  $f_d(x,x)=f(x)$,  
\item  $x_1 \geq x_2 \Rightarrow f_d(x_1,y)\geq f_d(x_2,y)$, and
\item $y_1 \geq y_2 \Rightarrow f_d(x,y_1) \leq f_d(x,y_2)$.
\end{enumerate} 
\end{defn}
\begin{prop}\cite[Theorem 1]{coogan2015efficient}\label{prop:embedding}
Let $f:\mathcal{X} \subseteq \mathbb{R}^n \rightarrow \mathcal{T} \subseteq \mathbb{R}^m$ be a mixed monotone mapping with decomposition function $f_d:\mathcal{X} \times \mathcal{X} \rightarrow \mathcal{T}$ and $\underline{x} \leq x \leq \overline{x}$, where $\underline{x},x,\overline{x} \in \mathcal{X}$. Then $f_d(\underline{x},\overline{x}) \leq f(x) \leq f_d(\overline{x},\underline{x})$.
\end{prop}
Note that the decomposition function of a 
vector field 
\yongn{is not unique and a specific one is given} 
in \cite[Theorem 2]{yang2019sufficient}\yongn{: If} a vector field $q=\begin{bmatrix} h^\top_1 & \dots & q^\top_n \end{bmatrix}^\top:X \subseteq \mathbb{R}^n \rightarrow \mathbb{R}^m$ is differentiable and its partial derivatives are bounded with known bounds, i.e., $\frac{\partial q_i}{\partial x_j} \in (a^q_{i,j},b^q_{i,j}), \forall x \in X \in \mathbb{R}^n$, where $a^q_{i,j},b^q_{i,j} \in \overline{\mathbb{R}}$, then $h$ is mixed monotone with a decomposition function $q_d=\begin{bmatrix} q^\top_{d1} & \dots & q^\top_{di} & \dots q^\top_{dn} \end{bmatrix}^\top$, where $q_{di}(x,y)=q_{i}(z)+(\alpha^q_i-\beta^q_i)^\top (x-y), \forall i \in \{1,\dots,n\}$, and $z,\alpha^q_i,\beta^h_i \in \mathbb{R}^n$ can be computed in terms of $x, y, a^q_{i,j}, b^q_{i,j}$ as given in \cite[(10)--(13)]{yang2019sufficient}. Consequently, for $x=[x_1\dots x_j \dots x_n]^\top$, $y=[y_1\dots y_j \dots y_n]^\top$, we have 
\begin{align}
q_{d}(x,y)=q(z)+C^q(x-y),  \label{eq:decompconstruct} 
\end{align}
where $C^q \hspace{-.1cm}  \triangleq \hspace{-.1cm}  \begin{bmatrix} [\alpha^q_1-\beta^q_1] & \hspace{-0.2cm}\dots \hspace{-0.2cm} & [\alpha^q_i-\beta^q_i] & \dots [\alpha^f_m-\beta^f_m] \end{bmatrix}^\top \hspace{-.2cm} \in \hspace{-.1cm}  \mathbb{R}^{m \times n}$, 
with 
$\alpha^f_i,\beta^f_i$ given in \cite[(10)--(13)]{yang2019sufficient}, $z=[z_1 \dots z_j \dots z_m]^\top$ and $z_j=x_j$ or $y_j$ (dependent on the case, cf. \cite[Theorem 1 and (10)--(13)]{yang2019sufficient} for details). On the other hand, when the precise lower and upper bounds, $a_{i,j}, b_{i,j}$, of the partial derivatives are not known or are hard to compute, we can obtain upper and lower approximations of the bounds by using Proposition \ref{prop:affine abstractions} 
with 
the slopes \yong{set} to zero. 
\section{Problem Formulation} \label{sec:Problem}
\noindent\textbf{\emph{System Assumptions.}} 
Consider the nonlinear discrete-time system with unknown inputs and bounded noise 
\begin{align} \label{eq:system}
\begin{array}{ll}
x_{k+1}&=f(x_k,d_k,u_k,w_k),\\
y_k&=g(x_k,d_k, u_k,v_k), 
\end{array} 
\end{align}
where $x_k \in \mathcal{X} \subset \mathbb{R}^n$ is the state vector at time $k \in \mathbb{N}$, 
$u_k \in \mathcal{U} \subset \mathbb{R}^m$ is a known input vector, $d_k \in \mathcal{D} \subset \mathbb{R}^p$ is an unknown dynamic input vector that its dynamics is governed by an \emph{unknown vector field} $h(.)$ as
\begin{align} \label{eq:input_dynamics}
d_{k+1}=h(x_k,d_k,u_k,w_k),
\end{align}
and $y_k \in \mathbb{R}^l$ is the measurement vector. The process noise $w_k \in \mathbb{R}^n$ and the measurement noise $v_k \in \mathbb{R}^l$ are assumed to be bounded, with $\underline{w} \leq w_k \leq \overline{w}$ and $\underline{v} \leq v_k \leq \overline{v}$, where $\underline{w}$, $\overline{w}$ and $\underline{v}$, $\overline{v}$ are the known \yongn{lower  and upper bounds of the process and measurement noise signals, respectively.} 
We also assume \yongn{that} lower and upper bounds, $\underline{z}_0$ and $\overline{z}_0$, for the initial augmented state ${z}_0 \triangleq \begin{bmatrix} x^\top_0 & d^\top_0 \end{bmatrix}^\top$ {are} available, i.e., $\underline{z}_0 \leq z_0 \leq \overline{z}_0$. The vector fields $f(\cdot):\mathbb{R}^n \times \mathbb{R}^p \times \mathbb{R}^m \times \mathbb{R}^n \rightarrow \mathbb{R}^n$ and $g(\cdot):\mathbb{R}^n \times \mathbb{R}^p \times \mathbb{R}^m \times \mathbb{R}^l \rightarrow \mathbb{R}^l$ are known, while 
the vector filed $h(\cdot)=\begin{bmatrix} h^\top_1(\cdot) \dots h^\top_p(\cdot) \end{bmatrix}^\top:\mathbb{R}^n \times \mathbb{R}^p \times \mathbb{R}^m \times \mathbb{R}^n  \rightarrow \mathbb{R}^p$ is \emph{unknown}, but each of its arguments $h_j(\cdot):\mathbb{R}^n \times \mathbb{R}^p \times \mathbb{R}^m \times \mathbb{R}^n  \rightarrow \mathbb{R}$, $\forall j \in \{1\dots p\}$ is known to be Lipschitz continuous with the known Lipschitz constant $L^h_j$.
Moreover, we assume 
the following: 
\begin{assumption}\label{assumption:mix-monotone}
Vector field $f(\cdot)$ is 
mixed-monotone with decomposition function $f_d(\cdot,\cdot):\mathbb{R}^n \times \mathbb{R}^p \times \mathbb{R}^m \times \mathbb{R}^n \times \mathbb{R}^n \times \mathbb{R}^p \times \mathbb{R}^m \times \mathbb{R}^n \rightarrow \mathbb{R}^n$. 
\end{assumption}
\begin{assumption} \label{assumption:state_boundedness}
The entire space $\mathbb{X} \triangleq \mathcal{Z} \times \mathcal{U}$ is bounded, where $\mathcal{Z} \triangleq \mathcal{X} \times \mathcal{D}$ and $\mathcal{U}$ are the spaces of the augmented states $z_k \triangleq \begin{bmatrix} x_k^\top & d_k^\top \end{bmatrix}^\top$ and the known inputs $u_k$, $\forall k \in \{0\dots \infty\}$, respectively. 
\end{assumption}
\yongn{Note} that 
Assumption \ref{assumption:mix-monotone} is satisfied for a broad range of nonlinear \yongn{functions \cite{yang2019tight}}, while Assumption \ref{assumption:state_boundedness} is reasonable for most practical systems.

The 
observer design problem 
 can be stated as follows:
\begin{problem}\label{prob:SISIO}
Given a partially known nonlinear discrete-time system \eqref{eq:system} with bounded noise signals and unknown dynamic inputs \eqref{eq:input_dynamics}, design a stable observer that simultaneously finds bounded intervals 
of compatible states and unknown inputs. 
\end{problem}

\section{State and Model Interval Observers (SMIO)} \label{sec:observer}
\subsection{Interval-Valued Recursive Observer} \label{sec:obsv}
A three-step recursive interval-valued observer that combines model-based and data-driven approaches will be considered in this paper. The observer structure is composed of a State Propagation (SP), a Measurement Update (MU) step and a Model Learning (ML) step. 
In the state propagation step, the interval for the augmented states (consisting of the state and the unknown input) is propagated for one time step through the nonlinear state equation and the upper and lower approximation of the unknown input function obtained in previous time step. 
In the update step, compatible intervals of the augmented states are iteratively updated given new measurements and observation function, and finally
 the model learning step re-estimates the upper and lower approximations (abstractions) for the function of the unknown inputs. More formally, the three observer steps have the following form (with $z_k\triangleq [x^\top_k \ d^\top_k]^\top$, $z^p_k\triangleq [x^{p\top}_k \ d^{p\top}_k]^\top$):     
\begin{align*}
&\text{\emph{SP:}}  \ \ \mathcal{I}^{z^p}_{k} = \mathcal{F}^p(\mathcal{I}^{z}_{k-1},y_{k-1},u_{k-1},\overline{h}_{k-1}(.),\underline{h}_{k-1}(.)),\\
&\text{\emph{MU:}}  \ \ \mathcal{I}^{z}_{k} = \mathcal{F}^u( \mathcal{I}^{z^p}_{k},y_{k},u_{k}),\\
&\text{\emph{ML:}}  [\underline{h}^\top_{k}(.) \ \overline{h}^\top_{k}(.)]^\top=  \mathcal{F}^l (\{\mathcal{I}^{z}_{k-t},u_{k-t}\}_{t=0}^{k}),
\end{align*}
with $\mathcal{F}^p$ and $\mathcal{F}^u$ being to-be-designed interval-valued mappings and $\mathcal{F}^l$ a to-be-constructed function over-approximation procedure (abstraction model), while $\mathcal{I}^{z^p}_{k}$ and $\mathcal{I}^z_{k}$ are the intervals of compatible propagated and estimated augmented states and $\{\overline{h}_k(\cdot),\underline{h}_k(\cdot)\}$ is a \emph{data-driven abstraction/over-approximation model} for the unknown function $h(\cdot)$, at time step $k$, respectively, i.e., $\forall \zeta_k \in \mathcal{D}_h: \underline{h}_k(\zeta_k) \leq h(\zeta_k) \leq \overline{h}_k(\zeta_k)$ at time step $k$, where $\mathcal{D}_h$ is the domain of $h(\cdot)$ and $\zeta_k\triangleq [z^\top_k \ u^\top_k \ w^\top_k]^\top$.

To leverage the properties of intervals  \cite{ellero2019unknown}, while taking to consideration the computational complexity of optimal observers \cite{milanese1991optimal}, we consider the following form of interval estimates in the propagation and update steps:
\begin{align*}
\mathcal{I}^{z^p}_{k}&=\{z \in \mathbb{R}^{n+p}: \underline{z}^p_{k} \leq z \leq \overline{z}^p_{k}\}, \\
\mathcal{I}^z_{k}&=\{z \in \mathbb{R}^{n+p}: \underline{z}_{k} \leq z \leq \overline{z}_{k}\}, 
\end{align*}
where the estimation boils down to find the maximal and minimal values of $\mathcal{I}^{z^p}_{k}$ and $\mathcal{I}^z_{k}$, i.e., $\overline{z}^p_{k},\underline{z}^p_{k},\overline{z}_{k},\underline{z}_{k}$. Further, at the model learning step, given the interval estimates for a certain period of time as data, we use a data-driven function abstraction/over-approximation model, developed in our previous work \cite{Jin2020datadriven}, to update our previously {estimated} model of the input dynamics $h(\cdot)$ in the current time step. 
\subsection{Observer's Structure}
Our interval observer can be defined at each time step $k \geq 1$ as follows (with augmented state $z_k \triangleq \begin{bmatrix} x^\top_k & d^\top_k \end{bmatrix}^\top$, $\zeta_k \triangleq \begin{bmatrix} z^\top_k & u^\top_k & w^\top_k \end{bmatrix}^\top$ and known $\underline{x}_{0}$ and $\overline{x}_0$ such that $\underline{x}_{0} \leq x_0 \leq \overline{x}_0$): \\
\noindent \textbf{\emph{State Propagation (SP)}}: \vspace{-0.1cm}
\begin{subequations}
\begin{align}
\label{eq:propagation}  &\hspace{-.1cm}\begin{bmatrix} \overline{x}^{p}_{k} \\ \underline{x}^{p}_{k} \end{bmatrix}  \hspace{-.15cm}=\hspace{-.15cm} \begin{bmatrix} \min(f_d(\overline{z}_{k-1},u_{k-1},\overline{w},\underline{z}_{k-1},u_{k-1},\underline{w}),\overline{x}^{a,p}_{k} ) \\ \max(f_d(\underline{z}_{k-1},u_{k-1},\underline{w},\overline{z}_{k-1},u_{k-1},\overline{w}),\underline{x}^{a,p}_{k})\hspace{-.1cm} \end{bmatrix}\hspace{-.1cm}, \\
   &\hspace{-.1cm}\begin{bmatrix} \overline{d}^{p}_{k} \\ \underline{d}^{p}_{k} \end{bmatrix} \hspace{-.1cm}=\hspace{-.1cm}\mathbb{A}^h_k\begin{bmatrix} \overline{z}^p_{k-1} \\ \underline{z}^p_{k-1} \end{bmatrix}\hspace{-.1cm}+\hspace{-.1cm} \mathbb{B}^h_k u_{k-1}+\mathbb{W}^h_k \begin{bmatrix} \overline{w} \\ \underline{w} \end{bmatrix}+\tilde{e}^h_k,\label{eq:d_prop}\\ 
 & \overline{z}^p_{k}  \hspace{-.1cm}=\hspace{-.1cm}\begin{bmatrix} \overline{x}^{p^\top}_{k} & \overline{d}^{p^\top}_{k}  \end{bmatrix}^\top,      \underline{z}^p_{k}=\begin{bmatrix} \underline{x}^{p^\top}_{k} & \underline{d}^{p^\top}_{k}  \end{bmatrix}^\top, \label{eq:x_prop_up}
\end{align}
\end{subequations}
\noindent \textbf{\emph{Measurement Update (MU)}}: \vspace{-0.1cm}
\begin{subequations}
\begin{align}
&\hspace{-2cm} \begin{bmatrix} \overline{z}_{k} &\underline{z}_{k} \end{bmatrix}= \lim_{i \to \infty} \begin{bmatrix} \overline{z}^{u}_{i,k} & \underline{z}^{u}_{i,k} \end{bmatrix}, \label{eq:mup} \\
  &\hspace{-2cm}  \begin{bmatrix} \overline{x}_k & \underline{x}_k \\ \overline{d}_k &  \underline{d}_k \end{bmatrix}=\begin{bmatrix} \overline{z}_{k,(1:n)} & \underline{z}_{k,(1:n)} \\ \overline{z}_{k,(n+1:n+p)} & \underline{z}_{k,(n+1:n+p)} \end{bmatrix},  
 \end{align}
 \end{subequations}
\noindent \textbf{\emph{Model Learning (ML)}}:  \vspace{-0.1cm}
\begin{subequations}
\begin{align}
& \overline{h}_{k,j}(\zeta_k) \hspace{-0.05cm}=  \min\limits_{t \in \{T-1,\hdots, 0\}}  (\overline{d}_{k-t,j}\hspace{-0.1cm}+\hspace{-0.1cm}L^{h}_j\|\zeta_k\hspace{-0.1cm}-\hspace{-0.1cm}\tilde{\zeta}_{k-t}\|)\hspace{-0.1cm}+\hspace{-0.1cm}\varepsilon^j_{k-t} \hspace{-0.05cm}, \label{upper_func}\\[-0.2em]
& \underline{h}_{k,j}(\zeta_k) \hspace{-0.05cm}=  \max_{\substack{t \in \{T-1,\hdots, 0\}}}  (\underline{d}_{k-t,j}\hspace{-0.1cm}-\hspace{-0.1cm}L^{h}_j\|\zeta_k\hspace{-0.1cm}-\hspace{-0.1cm}\tilde{\zeta}_{k-t}\|)\hspace{-0.1cm}+\hspace{-0.1cm}\varepsilon^j_{k-t} \hspace{-0.05cm}, \label{lower_func}
 \end{align}
 \end{subequations}
where $j \in \{1\dots p\}$, $\{\tilde{\zeta}_{k-t}=(1/2)(\overline{\zeta}_{k-t}+\underline{\zeta}_{k-t})\}_{t=0}^{k}$ and $\{\overline{d}_{k-t},\underline{d}_{k-t}\}_{t=0}^{k}$ are the \emph{augmented} input-output data set. At each time step $k$, the augmented data set constructed from the estimated framers gathered from the initial to the current time step, is used in the model learning step to recursively derive over-approximations of the unknown function $h(\cdot)$, i.e., $\{\overline{h}_k(.),\underline{h}_k(.)\}$ 
 by applying \cite[Theorem 1]{Jin2020datadriven}.  
 In addition
\begin{align}
  \begin{bmatrix} \overline{x}^{a,p}_{k} \\ \underline{x}^{a,p}_{k} \end{bmatrix} \ =\mathbb{A}^f_k\begin{bmatrix} \overline{z}^p_{k-1} \\ \underline{z}^p_{k-1} \end{bmatrix}\hspace{-.1cm}+\hspace{-.1cm}\mathbb{B}^f_k u_{k-1}+\mathbb{W}^f_k \begin{bmatrix} \overline{w} \\ \underline{w}\end{bmatrix}+\tilde{e}^f_k \label{eq:x_prop_abst} .
\end{align}
Moreover, the \emph{sequences of updated framers} $\{\overline{z}^u_{i,k},\underline{z}^u_{i,k}\}_{i=1}^{\infty}$ are iteratively computed  as follows:  
\begin{align}
\label{eq:zupp}&\begin{bmatrix} \overline{z}^u_{0,k} & \underline{z}^u_{0,k} \end{bmatrix} = \begin{bmatrix} \overline{z}^p_k & \underline{z}^p_k \end{bmatrix},  \quad  \forall i \in \{1\dots \infty \}: \\
&\begin{bmatrix}\overline{z}^{u}_{i,k} \\ \underline{z}^{u}_{i,k} \end{bmatrix}\hspace{-.1cm}=\hspace{-.1cm}\begin{bmatrix} \min(A^{g\dagger +}_{i,k} \overline{\alpha}_{i,k}\hspace{-.1cm}-\hspace{-.1cm}A^{g\dagger ++}_{i,k} \underline{\alpha}_{i,k}\hspace{-.1cm}+\hspace{-.1cm}\omega_{i,k},\overline{z}^{u}_{i-1,k}) \\ \max(A^{g\dagger +}_{i,k} \underline{\alpha}_{i,k}\hspace{-.1cm}-\hspace{-.1cm}A^{g\dagger ++}_{i,k} \overline{\alpha}_{i,k}\hspace{-.1cm}-\hspace{-.1cm}\omega_{i,k},\underline{z}^{u}_{i-1,k}) \end{bmatrix}\hspace{-.1cm}, \label{eq:iter_update}
\end{align} 
where
\begin{align}
&\begin{bmatrix} \overline{t}_{i,k} \\ \underline{t}_{i,k} \end{bmatrix}\hspace{-.15cm}=\hspace{-.15cm}\begin{bmatrix} y_k-B^{g}_{i,k}u_k \\ y_k-B^{g}_{i,k}u_k \end{bmatrix} \hspace{-.15cm}+\hspace{-.15cm}\begin{bmatrix} W^{g++}_{i,k} & -W^{g+}_{i,k} \\ \ -W^{g+}_{i,k} & W^{g++}_{i,k} \end{bmatrix}\hspace{-.15cm}  \begin{bmatrix}\overline{v} \\ \underline{v}\end{bmatrix}\hspace{-.15cm}-\hspace{-.15cm}\begin{bmatrix}\underline{e}^{g}_{i,k} \\ \overline{e}^{g}_{i,k} \end{bmatrix} \label{eq:t} \\
&\begin{bmatrix}\overline{\alpha}_{i,k} \\ \underline{\alpha}_{i,k} \end{bmatrix}=\begin{bmatrix} \min(\overline{t}_{i,k},A^{g+}_{i,k}\overline{z}^u_{i-1,k}-A^{g++}_{i,k}\underline{z}^u_{i-1,k}) \\ \max(\underline{t}_{i,k},A^{g+}_{i,k}\underline{z}^u_{i-1,k}-A^{g++}_{i,k}\overline{z}^u_{i-1,k}) \end{bmatrix}. \label{eq:alpha}
\end{align}
Furthermore, $\forall q \in \{f,h\}, \mathbb{J} \in \{\mathbb{A},\mathbb{W}\}$, $ i \in \{1\dots \infty \}$, $j \in \{1\dots p\}$, $\omega_{i,k}$, $A^g_{i,k}$, $B^g_{i,k}$, $W^g_{i,k}$, $\overline{e}^g_{i,k}$, $\underline{e}^g_{i,k}$, $\mathbb{B}^q_{k}$, $\mathbb{J}^q_{k}$, $\tilde{e}^q_{k}$, $\varepsilon^j_{k-t}$ and $f_{d}(.,.,.,.)$ 
  are to-be-designed observer parameters, matrix gains (with appropriate dimensions) and bounding function, at time $k$ and iteration $i$, with the purpose of achieving desirable observer properties. 
  
  Note that since the tightness of the upper and lower bounding functions for the observation function $g$ (cf. Propositions \ref{prop:affine abstractions} and \ref{prop:embedding}) depends on the \emph{a priori} interval $\mathcal{B}$, the measurement update step is done iteratively (see proof of Theorem \ref{thm:stability} for more explanation).  
Hence, if tighter updated intervals are obtained starting from the compatible intervals from the propagation step, we can use them as the new $\mathcal{B}$ to obtain better abstraction/bounding functions for $g$, which in turn may lead to even tighter updated intervals. Repeating this process results in a sequence of monotonically tighter updated intervals, that is convergent by the monotone convergence theorem, and its limit is chosen as the final interval estimate at time $k$. 

Further, benefiting from our previous result in \cite[Theorem 1]{Jin2020datadriven}, where we developed a data-driven approach for over-approximation/abstraction of Lipschitz unknown nonlinear functions given noisy data, in the model learning step, we treat the history of obtained compatible intervals in the past time steps up to the current time, $\{[\underline{z}_s,\overline{z}_s]\}_{s=0}^{k}$ as the noisy input data and the compatible interval of unknown inputs, $[\underline{d}_k,\overline{d}_k]$, as the noisy output data to recursively construct a sequence of \emph{abstraction/over-approximation models} \{$\overline{h}_k(\cdot),\underline{h}_k(\cdot)\}_{k=1}^{\infty}$ for the unknown input function $h(\cdot)$, that by construction satisfy \eqref{eq:tight_estimate}, i.e. our input model estimation is correct and becomes more precise over time (cf. Lemma \ref{lem:tight_model}). 
  Algorithm \ref{algorithm1} summarizes the SMIO observer.
  \begin{algorithm}[t] \small
\caption{SMIO}\label{algorithm1}
\begin{algorithmic}[1]
		\State Initialize: $\text{maximal}(\mathcal{I}^z_0)=\overline{z}_0$; $\text{minimal}(\mathcal{I}^z_0)=\underline{z}_0$;
		\LineComment{\hspace{-0.05cm}Observer Gains Computation}
		\Statex  \hspace{-0.05cm}$\forall q \in \{f,h\}, \mathbb{J} \in \{\mathbb{A},\mathbb{W}\}$, $ i \in \{1\dots \infty \}$, $j \in \{1\dots p\}$ compute $\omega_{i,k}$, $A^g_{i,k}$, $B^g_{i,k}$, $W^g_{i,k}$, $\overline{e}^g_{i,k}$, $\underline{e}^g_{i,k}$, $\mathbb{B}^q_{k}$, $\mathbb{J}^q_{k}$, $\tilde{e}^q_{k}$, $\varepsilon^j_{k-t}$ \hspace{-0.1cm} via \hspace{-0.05cm}Theorem \ref{thm:framer} and Appendix \ref{subsec:thmproof} ;
				\For {$k =1$ to $\overline{K}$}

    	\LineComment{\hspace{-0.05cm}Augmented State Estimation}
		\Statex \hspace{0.1cm} Compute $\overline{z}^p_{k},\underline{z}^p_{k}$ via\hspace{-0.05cm} \eqref{eq:propagation}--\eqref{eq:x_prop_up}\hspace{-0.05cm} and $\{\overline{z}^{u}_{i,k},\underline{z}^{u}_{i,k}\}_{i=0}^{\infty}$\hspace{-0.05cm} via \eqref{eq:zupp}--\eqref{eq:alpha};
		
		\State \hspace{-0.3cm}$(\overline{z}_k,\underline{z}_k)=(\overline{z}^{u}_{\infty,k},\underline{z}^{u}_{\infty,k})$; $\mathcal{I}^z_{k}\hspace{-.1cm}=\hspace{-.1cm}\{z \in \mathbb{R}^n : \underline{z}_k\hspace{-.1cm} \leq z\hspace{-.1cm} \leq \overline{z}_{k}\}$;
				\Statex \hspace{0.1cm} Compute 
		$\delta^z_{k}$ through Lemma \ref{lem:convergence};
		
		\LineComment{\hspace{-0.05cm}Model Estimation}
		 \Statex \hspace{0.1cm} 
		 Compute $\overline{h}_{k}(\cdot),\underline{h}_{k}(\cdot)$ via \eqref{upper_func}--\eqref{lower_func}; 
		\EndFor

		\end{algorithmic}

\end{algorithm}   
\vspace{-.2cm}   
 \subsection{Correctness of the Observer}
 The objective of this section is to design the SMIO observer's gains such that the \emph{framer property} \cite{mazenc2013robust} holds, i.e., we desire to guarantee that the observer returns correct interval estimates, in the sense that 
starting from the initial interval $\underline{z}_0 \leq z_0 \leq \underline{z}_0$, the true augmented states of the dynamic system \eqref{eq:system} are guaranteed to be within the estimated intervals, 
given by \eqref{eq:propagation}-\eqref{lower_func}. 
If the observer is correct, we call $\{\overline{z}_k,\underline{z}_k\}_{k=0}^{\infty}$ 
an \emph{augmented state framer sequence} for system \eqref{eq:system}. 

Before deriving our main first result on correctness of the observer, we state a modified version of our previous result in \cite[Theorem 1]{singh2018mesh}, in a unified manner that enables us to derive parallel global and local affine bounding functions for our known $f(\cdot),g(\cdot)$ and unknown $h(\cdot)$ vector fields.    
\begin{prop}[Parallel Afine Abstarctions]\label{prop:affine abstractions}
Let the entire space be defined as $\mathbb{X}$ and suppose that Assumption \ref{assumption:state_boundedness} holds. Consider the vector fields $\overline{q}(.),\underline{q}(.):\mathbb{X} \subset \mathbb{R}^{n'} \to \mathbb{R}^{m'}$, where $ \forall \zeta \in \mathbb{X},\underline{q}( \zeta) \leq \overline{q}( \zeta)$, along with the 
following Linear Program (LP): 
\begin{subequations}  
\begin{align} 
\label{eq:abstraction} &\min\limits_{\theta^q_\mathcal{B},{A}^q_\mathcal{B},\overline{e}^q_\mathcal{B},\underline{e}^q_\mathcal{B}} {\theta^q_\mathcal{B}} \\
\nonumber  \ &s.t \ \ {A}^q_\mathcal{B} { \zeta}_{s}+\underline{e}^q_\mathcal{B}+\sigma^q \leq \underline{q}({ \zeta}_{s}) \leq  \overline{q}({x}_{s}) \leq {A}^q_\mathcal{B} { \zeta}_{s}+\overline{e} ^q_\mathcal{B}-\sigma^q, \\
\nonumber &\quad \quad  \overline{e}_{\mathcal{B}}^q-\underline{e}_{\mathcal{B}}^q-2\sigma^q \leq \theta^q \mathbf{1}_{m'} , \forall  \zeta_s \in \mathcal{V}_{\mathcal{B}},\\
 & \quad \quad  \underline{e}^q-\underline{e}^q_\mathcal{B} \leq ({A}^q_\mathcal{B}-\mathbb{A}^q) \zeta_{s} \leq \overline{e}^q-\overline{e}^q_\mathcal{B}, \label{eq:guarantee}
\end{align}  
\end{subequations}
where $\mathcal{B}$ is an interval with $\overline{\zeta},\underline{\zeta}$ and $\mathcal{V}_{\mathcal{B}}$ being its maximal, minimal and set of vertices, respectively, $\mathbf{1}_{m'} \in \mathbb{R}^m$ is a vector of ones, $\sigma^q$ is given in \cite[Proposition 1 and (8)]{singh2018mesh} for different classes of vector fieilds 
and $(\mathbb{A}^q,\overline{e}^q,\underline{e}^q)$ is the global parallel affine abstraction matrices for the pair of functions $\overline{q}(.),\underline{q}(.)$ on the entire space $\mathbb{X}$, i.e., 
\begin{align} \label{eq:global_abs}
\mathbb{A}^q{ \zeta}+\underline{e}^q \leq \underline{q}( \zeta) \leq \overline{q}(\zeta) \leq  \mathbb{A}^q { \zeta}+\overline{e}^q, \forall  \zeta \in \mathbb{X}.
\end{align} 

Suppose that $(\mathbb{A}^q,\overline{e}^q,\underline{e}^q)$ are not known. Then, solving \eqref{eq:abstraction} 
on the entire space $\mathbb{X}$, i.e., when $\mathcal{B}=\mathbb{X}$ (where the constraint \eqref{eq:guarantee} is trivially satisfied and is thus redundant) returns a tuple of $(\theta^q,\mathbb{A}^q,\overline{e}^q,\underline{e}^q)$ that satisfies \eqref{eq:global_abs}, i.e. constructs a global affine abstraction model for the pair of functions $\overline{q}(.),\underline{q}(.)$ on the entire space $\mathbb{X}$.  

Now, suppose that $(\mathbb{A}^q,\overline{e}^q,\underline{e}^q)$ are known (or have been computed as described above). Then, solving \eqref{eq:abstraction} 
on $\mathcal{B}$ constrained to \eqref{eq:guarantee}, returns a tuple of local parallel affine abstraction matrices for the pair of functions $\{\underline{q}(\cdot),\overline{q}(\cdot)\}$ on the interval $\mathcal{B}$, satisfying the following: $\forall  \zeta \in \mathcal{B}$
\begin{align} \label{eq:local_abs}
\mathbb{A}^q{ \zeta}\hspace{-.1cm}+\hspace{-.1cm}\underline{e}^q \hspace{-.1cm}\leq \hspace{-.05cm}{A}^q_\mathcal{B}{ \zeta}\hspace{-.1cm}+\hspace{-.1cm}\underline{e}^q_\mathcal{B} \hspace{-.05cm}\leq \hspace{-.05cm}\underline{q}( \zeta) \hspace{-.05cm}\leq\hspace{-.05cm} \overline{q}(\zeta) \hspace{-.1cm}\leq \hspace{-.1cm} {A}^q_\mathcal{B} { \zeta}\hspace{-.1cm}+\hspace{-.1cm}\overline{e}^q_\mathcal{B} \hspace{-.1cm} \leq \hspace{-.1cm} \mathbb{A}^q { \zeta}\hspace{-.1cm}+\hspace{-.1cm}\overline{e}^q\hspace{-.1cm}.
\end{align} 
\end{prop}
Now, equipped with all the required tools, we state our first main result on the framer property of the SMIO observer. 
\begin{thm}[Correctness of the Observer] \label{thm:framer}
Consider the system \eqref{eq:system} with its augmented state defined as $z \triangleq \begin{bmatrix} x^\top & d^\top \end{bmatrix}^\top$, along with the SMIO observer in \eqref{eq:propagation}--\eqref{lower_func}. Suppose that Assumptions \ref{assumption:mix-monotone}--\ref{assumption:state_boundedness} hold, 
$f_d(\cdot)$ is a decomposition function of $f(\cdot)$ and observer gains and parameters are designed as given in \eqref{subsec:thmproof}. Then, the SMIO observer is correct, i.e., the sequences $\{\overline{z}_k,\underline{z}_k\}_{k=0}^\infty$ construct framers for the augmented state sequence of system \eqref{eq:system}.
\end{thm}
Next, we show that given correct interval estimates, the abstraction model of the unknown input function becomes tighter (i.e., more precise) over time, so our estimation of the unknown input model becomes more accurate over time. 
\begin{lem}\label{lem:tight_model}
Consider the system \eqref{eq:system} and the SMIO observer in \eqref{eq:propagation}--\eqref{lower_func} and suppose that all the assumptions in Theorem \ref{thm:framer} hold. Then, the following holds: 
\begin{align}\label{eq:tight_estimate}
\begin{array}{ll}
\underline{h}_0(\zeta_0)\hspace{-.1cm} \leq\hspace{-.1cm} \dots\hspace{-.1cm} \leq\hspace{-.1cm} \underline{h}_k(\zeta_k)\hspace{-.1cm} \leq \hspace{-.1cm}\dots\hspace{-.1cm} \leq \lim_{k \to \infty}\underline{h}_{k}(\zeta_k) \hspace{-.1cm}\leq\hspace{-.1cm} {h}(\zeta_k)\\
{h}(\zeta_k) \hspace{-.1cm}\leq \lim_{k \to \infty}\overline{h}_{k}(\zeta_k)\hspace{-.1cm} \leq \hspace{-.1cm}\dots\hspace{-.1cm} \leq \hspace{-.1cm}\overline{h}_k(\zeta_k) \leq \hspace{-.1cm}\dots\hspace{-.1cm} \hspace{-.1cm}\leq \hspace{-.1cm}\overline{h}_0(\zeta_0),
\end{array}
\end{align}
i.e, the unknown input model estimations/abstractions are correct and become more precise/tighter in time. 
\end{lem}
\subsection{ Observer Stability}
In this section, we study the stability of the designed observer. We first formally define the notion of stability that we investigate in this paper.
\begin{defn}[Stability] \label{defn:stability}
The observer SMIO \eqref{eq:propagation}-\eqref{lower_func} is stable, if the interval widths sequence $\{\|\Delta^z_{k-1}\| \triangleq \|\overline{z}_{k-1}-\underline{z}_{k-1}\|\}_{k=1}^{\infty}$ is {uniformly bounded}, and consequently the sequence of estimation errors $\{\|\tilde{z}_{k-1}\| \triangleq \max (\|z_{k-1}-\underline{z}_{k-1}\|,\|\overline{z}_{k-1}-{z}_{k-1}\|)$ is also uniformly bounded.
\end{defn}
Next, we derive a property for the decomposition function given in \eqref{eq:decompconstruct}, which will be helpful in deriving sufficient conditions for the observer's stability.
\begin{lem}\label{lem:lip-dec}
Let $q(\zeta):\mathbb{X} \subset \mathbb{R}^n \to \mathbb{R}^m$ be a mixed-monotone vector-field with a corresponding decomposition function $q_d(.,.)$ constructed using \eqref{eq:decompconstruct}. Suppose that Assumption \ref{assumption:state_boundedness} holds and let $(\mathbb{A}^q,\overline{e}^q,\underline{e}^q)$ be the parallel affine abstraction matrices for function $q(\cdot)$ on its entire domain $\mathbb{X}$ (can be computed via Proposition \ref{prop:affine abstractions}). Consider any ordered pair $\underline{\zeta} \leq \overline{\zeta} \in \mathbb{X}$. Then, $ \Delta q_{\zeta} \leq (|\mathbb{A}^q|+2C^q) \Delta \zeta+\Delta e^q$,
  with $\Delta q_{\zeta} \triangleq {q}_d(\overline{\zeta},\underline{\zeta})-{q}_d(\underline{\zeta},\overline{\zeta})$, $\Delta \zeta \triangleq \overline{\zeta}-\underline{\zeta}$ and $C^q$ given in \eqref{eq:decompconstruct}.
\end{lem}
Now we are ready to state our next main result on the SMIO observer's stability through the following theorem. 
\begin{thm}[
Observer Stability] \label{thm:stability}
Consider the system \eqref{eq:system} along with the SMIO observer in \eqref{eq:propagation}--\eqref{lower_func}. Let $\mathbb{D}_{n'}$ be the set of all diagonal matrices in $\mathbb{R}^{n' \times n'}$ with their diagonal arguments being $0$ or $1$. Suppose that all the assumptions in Theorem \ref{thm:framer} hold and the decomposition function $f_d$ is constructed using \eqref{eq:decompconstruct}. 
Then, the observer is stable 
if 
\begin{subequations}
\begin{align}
\mathcal{L}^* \triangleq \min\limits_{(D_1,D_2,D_3) \in \mathbb{D}^*} &\| \mathcal{A}^g(D_1,D_2)\mathcal{A}^{f,h}(D_3)\|  \leq 1, \label{eq:stability} \\
  s.t. &\ \  D_{1,i,i}=0 \ \text{if} \ r_i=1, \label{eq:stab_constr}
\end{align}
\end{subequations} 
with $\mathcal{A}^g(D_1,D_2)  \triangleq (I-D_1)+D_1|A^{g\dagger}|(I-D_2)|A^g|$, $\mathcal{A}^{f,h}(D_3)\triangleq\begin{bmatrix} (|A^f|+2(I-D_3)C^f_z)^\top & |A^h|^\top \end{bmatrix}^\top$, $\{A^q \triangleq \mathbb{A}^q_{(1:n+p)}\}_{q \in \{f,g,h\}}$, $\mathbb{A}^q$ given in Proposition \ref{assumption:state_boundedness}, $r \triangleq \textstyle{rowsupp}(I-A^{g\dagger}A^g)$, $C^f \triangleq \begin{bmatrix} C^f_z & C^f_u & C^f_w \end{bmatrix}$ given in \eqref{eq:decompconstruct} and $\mathbb{D}^* \triangleq \mathbb{D}_{n+p} \times \mathbb{D}_l \times \mathbb{D}_n $.
\end{thm}
\begin{rem}
The optimization problem in \eqref{eq:stability}--\eqref{eq:stab_constr} is a (mixed-)integer program with a finitely countable feasible set 
($|\mathbb{D}^*| \leq 2^{2n+p+l}$), which can be easily solved by enumerating all possible solutions and comparing their values.  
\end{rem}
We conclude this section by providing upper bounds for the interval widths and compute their \yong{steady-state values, if they exist}.}
\begin{lem}[Upper Bounds of the Interval Widths and their Convergence]\label{lem:convergence}
Consider the system \eqref{eq:system} and the observer \eqref{eq:propagation}--\eqref{lower_func} and suppose all the assumptions in Theorem \ref{thm:stability} hold. Then, the sequence of $\{\Delta^z_k \triangleq \overline{z}_k-\underline{z}_k\}_{k=0}^{\infty}$ is uniformly upper bounded by a convergent sequence as: 
\begin{align} \label{eq:error_upper}
\Delta^z_k &\leq \overline{\mathcal{A}}^{k}\Delta^z_0+\sum_{j=0}^{k-1} \overline{\mathcal{A}}^j\overline{\Delta} \xrightarrow{k \to \infty} e^{\overline{\mathcal{A}}} \overline{\Delta},
\end{align}
where 
 \begin{align}
\nonumber &\hspace{-.15cm}\overline{\mathcal{A}} =\mathcal{A}(D^*_1,D^*_2,D^*_3) \triangleq \mathcal{A}^g(D^*_1,D^*_2)\mathcal{A}^{f,h}(D^*_3), \\
\nonumber &\hspace{-.15cm}\overline{\Delta} =\Delta^g(D^* _1,D^*_2)+\mathcal{A}^g(D^*_1,D^*_2) \Delta^{f,h}(D^*_3), \\
\nonumber &\hspace{-.15cm}\mathcal{A}^g(D_1,D_2) \triangleq D_1|A^{g\dagger}|D_2 |A^g|+(I-D_1), \\
\nonumber &\hspace{-.15cm}\mathcal{A}^{f,h}(D_3) \triangleq\begin{bmatrix} (|A^f|+2(I-D_3)C^f_z)^\top & |A^h|^\top \end{bmatrix}^\top, \\
\nonumber &\hspace{-.15cm}\Delta^g(D_1,D_2) \triangleq D_1|A^{g\dagger}|D_2(|W^g|\Delta v+\Delta e^g), \  \Delta^{f,h} (D_3)  \hspace{-.1cm} \triangleq  \\
\nonumber &\hspace{-.15cm} \begin{bmatrix} ((|W^f| \hspace{-.1cm}+ \hspace{-.1cm} 2(I \hspace{-.1cm}- \hspace{-.1cm}D_3)C^f_w) \Delta w \hspace{-.1cm}+ \hspace{-.1cm} \Delta^f_e)^\top  \hspace{-.1cm}&  \hspace{-.1cm}(|W^h|\Delta w \hspace{-.1cm} + \hspace{-.1cm} \Delta^h_e )^\top \end{bmatrix}^\top,
\end{align}
and $(D^*_1,D^*_2,D^*_3)$ is a solution of the following:
\begin{small}

\vspace{-.25cm}
\begin{align*}
&\min\limits_{D_1,D_2,D_3} \|e^{\mathcal{A}(D_1,D_2,D_3)}(\Delta^g(D _1,D_2)\hspace{-.05cm}+\hspace{-.05cm}\mathcal{A}^g(D_1,D_2) \Delta^{f,h}(D_3))\| \\
&\quad \text{s.t.}  (D_1,D_2,D_3) \hspace{-.05cm} \in  \hspace{-.05cm} \{ (D'_1,D'_2,D'_3)  \hspace{-.05cm} \in \hspace{-.05cm} \mathbb{D}^*  \vline \ \mathcal{L}^*  \hspace{-.05cm}< \hspace{-.05cm}1\ \& \ \eqref{eq:stab_constr}  \ \text{holds} \}. 
\end{align*}
\end{small}
Consequently, the sequence of interval widths $\{ \| \Delta^z_k \|\}_{k=1}^{\infty}$ is uniformly upper bounded by a convergent sequence as: 
\begin{align} \label{qe:width_upper}
\|\Delta^z_k\| \leq \delta^z_k \triangleq \| \overline{\mathcal{A}}^{k}\Delta^z_0+\sum_{j=0}^{k-1} \overline{\mathcal{A}}^j\overline{\Delta}\| \xrightarrow{k \to \infty} \|e^{\overline{\mathcal{A}}} \overline{\Delta}\|.
\end{align}
\end{lem}
\vspace{-0.15cm}
\section{Illustrative Example} \label{sec:examples}
\begin{figure}[t]
\begin{center}
\includegraphics[scale=0.158,trim=53mm 0mm 5mm 5mm,clip]{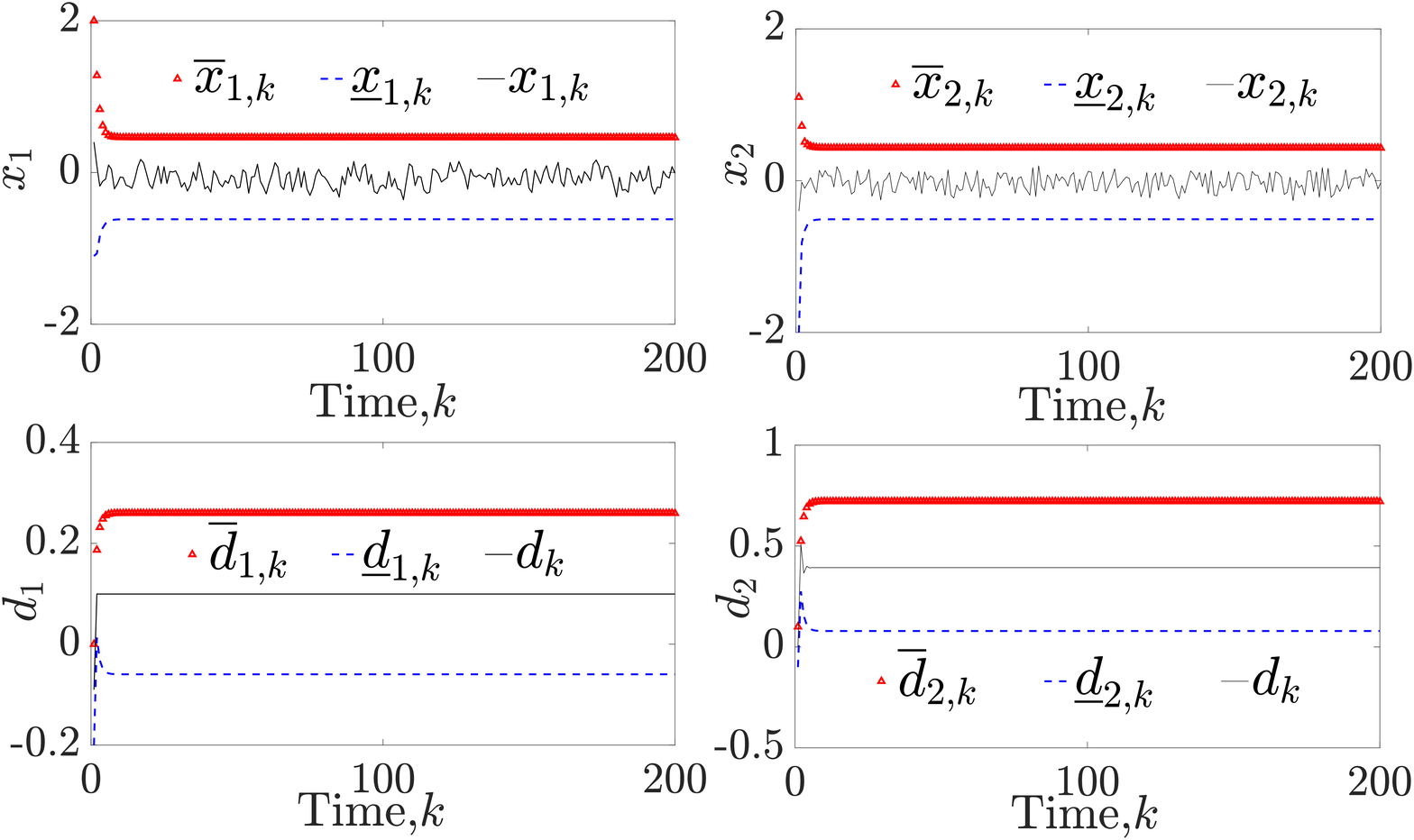}
\vspace{-0.15cm}
\caption{Actual states and inputs, $x_{1,k}$, $x_{2,k}$, $d_{1,k}$, $d_{2,k}$, as well as their estimated maximal and minimal values, $\overline{x}_{1,k}$, $\underline{x}_{1,k}$, $\overline{x}_{2,k}$, $\underline{x}_{1,k}$, $\overline{d}_{1,k}$, $\underline{d}_{1,k}$, $\overline{d}_{2,k}$, $\underline{d}_{2,k}$. \label{fig:variances2}}
\end{center}
\vspace{-0.25cm}
\end{figure}

We consider a slightly modified version of nonlinear dynamical system in \cite{de2019robust} with removing the uncertain parts of the matrices and including unknown dynamic inputs. The system can be described in the form \eqref{eq:system}--\eqref{eq:input_dynamics} with the following parameters:
$n=l=p=2$, $m=1$, $f(.)=\begin{bmatrix} f_1(.) & f_2(.)\end{bmatrix}^\top$, 
$g(.)=\begin{bmatrix} g_1(.) & g_2(.)\end{bmatrix}^\top$, 
$u_k=0$, 
$\overline{v}=-\underline{v}=\overline{w}=-\underline{w}=\begin{bmatrix} 0.2 & 0.2 \end{bmatrix}^\top$, $\overline{x}_0=\begin{bmatrix} 2 & 1.1 \end{bmatrix}^\top$, $\underline{x}_0=\begin{bmatrix} -1.1 & -2 \end{bmatrix}^\top$. Further,
\begin{align*}
\begin{array}{rl}
f_1(\zeta_k)&=0.6x_{1,k} - 0.12x_{2,k}+1.1 \sin(0.3x_{2,k}-0.2x_{1,k}),\\ f_2(\zeta_k)&=-0.2x_{1,k}-0.14x_{2,k},d_{1,k+1} = 0.1\cos(d_{1,k})\\
g_1(\nu_k)&=0.2x_{1,k}+0.65x_{2,k}+0.8\sin(0.3x_{1,k}+0.2x_{2,k}),\\
 g_2(\nu_k)&=\sin(x_{1,k}),d_{2,k+1} = \frac{1}{1+e^{d_{2,k}}}-0.1d_{1,k},
 \end{array}
\end{align*} 
with $\nu^\top_k \triangleq [z^\top_k \ u^\top_k \ v^\top_k]$. 
Moreover, using Proposition  \ref{prop:affine abstractions} while abstraction slopes are set to zero, we can obtain finite-valued upper and lower bounds (horizontal abstractions) for the partial derivatives of $f(\cdot)$ as: 
$\begin{bmatrix}a^f_{11} & a^f_{12} \\ a^f_{21} & a^f_{22}  \end{bmatrix}=\begin{bmatrix} 0.38 & -0.46 \\ -0.2-\epsilon & -0.14-\epsilon \end{bmatrix}$,    $\begin{bmatrix}b^f_{11} & b^f_{12} \\ b^f_{21} & b^f_{22}  \end{bmatrix}=\begin{bmatrix} 0.82 & 0.21 \\ -0.2+\epsilon & -0.14+\epsilon \end{bmatrix}$, 
where $\epsilon$ is a very small positive value, ensuring that the partial derivatives are in open intervals (cf. \cite[Theorem 1]{yang2019sufficient}). Therefore, Assumption \ref{assumption:mix-monotone} holds by \cite[Theorem 1]{yang2019sufficient}). 
Hence, we expect that the true states and unknown inputs are within the estimate intervals by Theorem \ref{thm:framer}, i.e., the interval estimates are correct. This can be observed from Figure \ref{fig:variances2}, where the true states and unknown inputs as well as interval estimates are depicted. 

\begin{figure}[t]
\begin{center}
\includegraphics[scale=0.147,trim=39mm 3mm 10mm 15mm,clip]{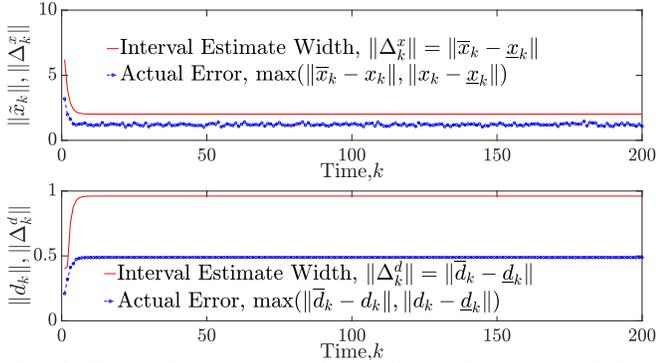}
\vspace{-0.1cm}
\caption{Estimation errors, estimate interval widths and their upper bounds for the interval-valued estimates of states, $\|\tilde{x}_{k|k}\|$, \hspace{-0.05cm}$\|\Delta ^x_k\|$, \hspace{-0.05cm}$\delta^x_k$, and unknown inputs, $\|\tilde{d}_{k}\|$, $\|\Delta ^d_k\|$, $\delta^d_k$. \label{fig:variances3}}
\end{center}
\vspace{-0.15cm}
\end{figure}

Furthermore, solving Proposition \ref{prop:affine abstractions} for global abstraction matrices, 
we derive $A^f=\begin{bmatrix} 0.4063  &  0.1706   &      0  & -0.1 &   1 &    0\\
    -0.2 & -0.14 &    0.2 &   -0.2 &  0 &   1 \end{bmatrix}$, $A^g=\begin{bmatrix} 0.4204  &  0.797 &  -0.1 &    0.3 &    1 &         0\\
         0.584 &  0  &     0.5 &   -0.7 &         0  &    1 \end{bmatrix}$, $A^h=\begin{bmatrix}0    &     0  & -0.0618  &       0   &      0  &       0\\
          0    &     0 &  -0.1669    &     0  &       0  &       0 \end{bmatrix}$ and from \cite[(10)--(13)]{yang2019sufficient}), we obtain $C_f=\begin{bmatrix}  0.374 & .02 \\ 0.0135 & 0.407\end{bmatrix}$, 
using \eqref{eq:decompconstruct}. 
Consequently, the mixed-integer program \eqref{eq:stability} constrained by \eqref{eq:stab_constr} 
results in $\mathcal{L}^*=1.1 >1$ and so 
the sufficient conditions in Theorem \ref{thm:stability} are not satisifed. 
Despite this, 
as can be seen in Figure \ref{fig:variances3}, we 
obtain {uniformly} bounded and convergent interval estimate errors when applying our observer design procedure, 
where at each time step, the actual error sequence is upper bounded by the interval widths, which converge to steady-state values.

\begin{figure}[t]
\begin{center}
\includegraphics[scale=0.138,trim=28mm 3mm 10mm 15mm,clip]{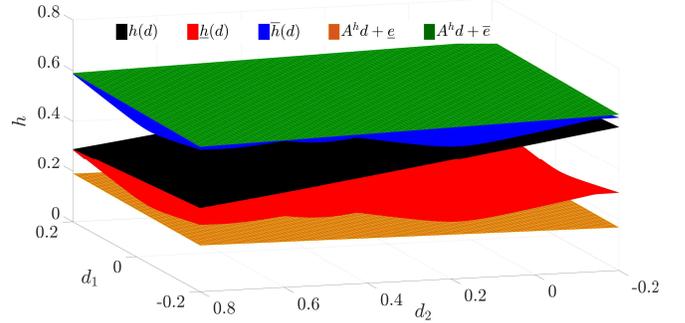}
\vspace{-0.1cm}
\caption{True input function $h(d_{k'})$, upper and lower local abstractions $\overline{h}_{k'}(d_{k'}),\underline{h}_{k'}(d_{k'})$ vs. global abstractions $A^hd_{k'}+\overline{e}^h,A^hd_{k'}\underline{e}^h$, at time step $k'=200$. \label{fig:abstraction}}
\end{center}
\vspace{-0.15cm}
\end{figure}

Note that as discussed in the proof of Theorem \ref{thm:stability}, since we need to check an \emph{a priori} condition (i.e., offline or before starting to implement the observer) for observer stability, we use global abstraction slopes for stability analysis. However, for the implementation, we iteratively update the framers and consequently, obtain the updated local abstractions, which, in turn, lead to updated local intervals that by construction are tighter than the global ones, as shown in the proof of Theorem \ref{thm:stability}. Hence, for a given system, it might be the case that the (relatively conservative) global-abstraction-based sufficient conditions for the observer stability given in Theorem \ref{thm:stability} do not hold, i.e., $\mathcal{L}^* >1$, while the implemented local-abstraction-based intervals are still uniformly bounded. This is the main benefit of using iterative local affine abstractions versus  global abstractions, with the cost of more extensive computational effort. Figure \ref{fig:abstraction} compares the tightness of intervals using global and iteratively updated local parallel affine abstractions.  
\vspace{-0.05cm}
\section{Conclusion} \label{sec:conclusion}
 An interval-valued observer for partially unknown nonlinear systems with dynamic unknown inputs and bounded noise signals was designed in this paper, that simultaneously estimated the augmented states and unknown input (with unknown  dynamics) of the system. By applying a combination of nonlinear bounding/decomposition functions and affine abstractions 
as well as benefiting from our previously developed data-driven function abstraction method to over-estimate the unknown input model from the noisy estimated input-output data, we showed that the estimate interval estimates are correct in the sense that 
our proposed observer recursively computes the maximal and minimal elements of the estimate intervals that are proven to contain the true augmented states, and by observing new output/measurement signals, iteratively shrinks the intervals by eliminating estimates that are not compatible with the measurements. 
Moreover, sufficient conditions for uniform boundedness of the sequence of estimate interval widths, i.e., stability of the designed observer were provided in the form of tractable  (mixed-)integer programs with finitely countable feasible sets.  
\bibliographystyle{unsrturl}

\bibliography{biblio}

\section*{Appendix: Observer Gain Definitions and Proofs} 
\subsection{Observer Gain Definitions}\label{subsec:thmproof}
$\forall \mathbb{J} \in \{\mathbb{A},\mathbb{W}\}, q \in \{f,h\}, J \in \{A,W\}, i \in \{1\dots \infty\}$: 
\begin{align*}
&\mathbb{J}^q_k\hspace{-.1cm}=\hspace{-.1cm}\begin{bmatrix} J^{q+}_{k} & -J^{q++}_{k} \\ -J^{q++}_{k} &J^{q+}_{k} \end{bmatrix}\hspace{-.1cm},\mathbb{B}^q_k\hspace{-.1cm}=\hspace{-.1cm}\begin{bmatrix} {B}^{q\top}_{k} & {B}^{q\top}_{k} \end{bmatrix}^\top\hspace{-.2cm},\tilde{e}^q_k\hspace{-.1cm}=\hspace{-.1cm}\begin{bmatrix} \overline{e}^{q\top}_{k} & \underline{e}^{q\top}_{k} \end{bmatrix}^\top\hspace{-.2cm},\\
&\omega_{i,k}\hspace{-.1cm}=\hspace{-.1cm}\kappa \textstyle{rowsupp}(I-A^{g\dagger}_{i,k}A^g_{i,k}), \varepsilon^j_{k-t}\hspace{-.1cm}=\hspace{-.1cm}2L^h_j\|\overline{\zeta}_{k-t}-\underline{\zeta}_{k-t}\|,
\end{align*}
$(A^g_{i,k}, B^g_{i,k},W^g_{i,k},\overline{e}^g_{i,k},\underline{e}^g_{i,k})$,$(A^q_{k}, B^q_{k},W^q_{k},\overline{e}^q_{k},\underline{e}^q_{k})_{q \in \{f,h\}}$
are solutions to the problem \eqref{eq:abstraction} 
for the corresponding functions $\{\underline{g}(\cdot)=\overline{g}(\cdot)=g(\cdot)\}$, $\{\underline{f}(\cdot)=\overline{f}(\cdot)=f(\cdot)\}$ and $\{\overline{h}_k(\cdot),\underline{h}_k(\cdot)\}$, on the intervals $[\begin{bmatrix} \underline{z}^{u\top}_{i-1,k} & u_{k-1}^\top & \underline{v}^\top  \end{bmatrix}^\top, \begin{bmatrix} \overline{z}^{u\top}_{i-1,k} & u_{k-1}^\top & \overline{v}^\top \end{bmatrix}^\top]$ for $g$ and $[\begin{bmatrix} \underline{z}_{k-1}^\top & u_{k-1}^\top & \underline{w}^\top  \end{bmatrix}, \begin{bmatrix} \overline{z}_{k-1}^\top & u_{k-1}^\top & \overline{w}^\top \end{bmatrix}^\top]$ for $f,\overline{h}_k,\underline{h}_k$, respectively, at time $k$ and iteration $i$, while 
 $\kappa$ is a very large positive real number (infinity). 
\subsection{Proof of Proposition \ref{prop:affine abstractions}}
Consider the case when the global affine abstraction matrices are unknown. Then, by setting $\mathcal{B}=\mathbb{X}$, $A^q_{\mathcal{B}}=\mathbb{A}^q$ and $\theta^q_{\mathcal{B}}$, constraint \eqref{eq:guarantee} is redundant and so the LP \eqref{eq:abstraction} boils down to a special case of the LP in \cite[(16)]{singh2018mesh}, with only one considered partition. Then, \eqref{eq:global_abs} follows from \cite[Theorem 1]{singh2018mesh}. Moreover, given the global affine abstractions, solving the LP in \eqref{eq:abstraction} is equivalent to solving the the LP in \cite[(16)]{singh2018mesh} on the corresponding interval (set) of $\mathcal{B}$, with the extra (non-trivial) constraint \eqref{eq:guarantee}.This constraint along with the result in \cite[Theorem 1]{singh2018mesh} result in \eqref{eq:local_abs}. 
 \qed
\subsection{Proof of Theorem \ref{thm:framer}}
To use induction and as for the induction base, by assumption, $\underline{z}_0 \leq z_0 \leq \overline{z}_0$ holds. Now for the induction step, suppose that $\underline{z}_{k-1} \leq z_{k-1} \leq \overline{z}_{k-1}$. Then, Propositions \ref{prop:bounding}--\ref{prop:affine abstractions} as well as \eqref{eq:system},\eqref{eq:x_prop_abst}--\eqref{eq:x_prop_up} and \cite[Theorem 1]{Jin2020datadriven} imply that $\underline{z}^p_{k} \leq z_{k} \leq \overline{z}^p_{k}$. Given this, iteratively obtaining upper and lower abstraction matrices for the observation function $g(.)$ based on  Proposition \ref{prop:affine abstractions} and applying Proposition \ref{prop:bounding}, result in 
\begin{align}\label{eq:aug_inequality}
\underline{\alpha}_{i,k} \leq A^g_{i,k}z_k \leq \overline{\alpha}_{i,k},
\end{align}
 where $\underline{\alpha}_{i,k},\overline{\alpha}_{i,k}$ are given in \eqref{eq:alpha} and $A^g_{i,k}$ is a solution of the LP in \eqref{eq:abstraction}, i.e., is a parallel abstraction slope for function $g(.)$ at iteration $i$ in the corresponding compatible interval $[\underline{z}^u_{i-1,k},\underline{z}^u_{i-1,k}]$. Then, multiplying \eqref{eq:aug_inequality} by $A^{g\dagger}_{i,k}$, Proposition \ref{prop:bounding}, the fact that $\underline{z}^u_{i-1,k},\overline{z}^u_{i-1,k}$ already construct framers for the augmented state $z_k$ at time $k$ and \cite{james1978generalised} imply that $\underline{z}^u_{i,k} \leq z_k \leq \overline{z}^u_{i,k}$, with $\underline{z}^u_{i,k},\overline{z}^u_{i,k}$ given in \eqref{eq:iter_update}.
 Now, note that by construction, the sequences of updated upper and lower framers,  $\{\overline{z}^u_{i,k}\}_{i=0}^{\infty}$ and $\{\underline{z}^u_{i,k}\}_{i=0}^{\infty}$ with $\overline{z}^u_{0,k}=\overline{z}^p_{k}$ and $\overline{z}^u_{0,k}=\overline{z}^p_{k}$, are monotonically decreasing and increasing, respectively, and hence are convergent by the \emph{monotone convergence theorem}. Consequently, their limits $\overline{z}_k,\underline{z}_k$ are the tightest possible framers, i.e., $\forall i \in \{1\dots \infty\}$:
 \begin{align*}
 \begin{array}{c}
   \underline{z}^{u}_{0,k} \leq \dots \leq \underline{z}^{u}_{i,k} \leq \dots  \leq  \lim_{i \to \infty}\underline{z}^{u}_{i,k}  \triangleq \underline{z}_k,\\ \overline{z}_k \triangleq \lim_{i \to \infty}\overline{z}^{u}_{i,k}   \leq \dots\leq \overline{z}^{u}_{i,k} \leq \dots \leq \overline{z}^u_{0,k}, 
  \end{array}
 \end{align*}
  where $\overline{z}_k,\underline{z}_k$ are the returned updated augmented state framers by the observer. This completes the proof. \qed 
   \subsection{Proof of Lemma \ref{lem:tight_model}}
 It directly follows from \cite[Theorem 1]{Jin2020datadriven} and Theorem \ref{thm:framer} that the model estimates are correct, i.e, $\forall k \in \{0\dots \infty\}: \underline{h}_k(\zeta_k) \leq h(\zeta_k) \leq  \overline{h}_k(\zeta_k)$. Moreover, considering the data-driven abstraction procedure in model learning step, note that by construction the data set used at time step $k$ is a subset of the one used at time $k+1$. Hence, by \cite[Proposition 2]{Jin2020datadriven} the abstraction model satisfies \emph{monotonicity}, i.e., \eqref{eq:tight_estimate} holds. \qed 
 \subsection{Proof of Lemma \ref{lem:lip-dec}}
 Starting form \eqref{eq:decompconstruct}, it is not hard to verify that 
 \begin{align} \label{eq:delta_q}
 \Delta q_{\zeta} = q(\zeta_1)-q(\zeta_2)+2C^q\Delta \zeta,
 \end{align}
 for some $\zeta_1,\zeta_2$ that satisfy $\underline{\zeta} \leq \zeta_1,\zeta_2 \leq \overline{\zeta}$. On the other hand, by Proposition \ref{prop:affine abstractions} in addition to Proposition \ref{prop:bounding}, $\forall j \in \{1,2\}$:
 \begin{align*}
 \mathbb{A}^{q+}\underline{\zeta}-\mathbb{A}^{q++}\overline{\zeta}+\underline{e}^q \leq q(\zeta_j) \leq \mathbb{A}^{q+}\overline{\zeta}-\mathbb{A}^{q++}\underline{\zeta}+\overline{e}^q,
 \end{align*}
 which implies $q(\zeta_1)-q(\zeta_2) \leq |\mathbb{A}^q|\Delta q_{\zeta}+\Delta e^q$. Combining this and \eqref{eq:delta_q} 
  yields the result.  \qed
\subsection{Proof of Theorem \ref{thm:stability}}  
Note that our goal is to obtain sufficient stability conditions that can be checked \emph{a priori} instead of for each time step $k$. On the other hand, for the implementation of the update step, we iteratively find new \emph{local} parallel abstraction slopes $A^g_{i,k}$ by iteratively solving the LP \eqref{eq:abstraction} for $g$ on the intervals obtained in the previous iteration, $\mathcal{B}^{u}_{i,k}=[\underline{z}^{u}_{i-1,k},\overline{z}^{u}_{i-1,k}]$, to find \emph{local} framers $\overline{z}^{u}_{i,k},\underline{z}^{u}_{i,k}$ (cf. \eqref{eq:zupp}--\eqref{eq:alpha}), with additional 
constraints given in \eqref{eq:guarantee} in the optimization problems, which guarantees that the iteratively updated \emph{local} intervals obtained using the local abstraction slopes are inside the global interval, i.e.,
\begin{align*}
 \begin{array}{c}
  \underline{z}^u_k \leq \underline{z}^{u}_{0,k} \leq \dots \leq \underline{z}^{u}_{i,k} \leq \dots  \leq  \lim_{i \to \infty}\underline{z}^{u}_{i,k}  \triangleq \underline{z}_k,\\ \overline{z}_k \triangleq \lim_{i \to \infty}\overline{z}^{u}_{i,k}   \leq \dots\leq \overline{z}^{u}_{i,k} \leq \dots \leq \overline{z}^u_{0,k} \leq \overline{z}^u_k, 
  \end{array}
\end{align*}
 where $\overline{z}^u_k,\underline{z}^u_k$ can be obtained by applying \eqref{eq:iter_update} for just one iteration (dropping index $i$) while $\overline{z}^u_{k,0}=\overline{z}^p_k,\underline{z}^u_{k,0}=\underline{z}^p_k$, as:  
 \begin{align}
 &\begin{bmatrix}\overline{z}^{u}_{k} \\ \underline{z}^{u}_{k} \end{bmatrix}\hspace{-.1cm}=\hspace{-.1cm}\begin{bmatrix} \min(A^{g\dagger +} \overline{\alpha}_k\hspace{-.1cm}-\hspace{-.1cm}A^{g\dagger ++} \underline{\alpha}_k\hspace{-.1cm}+\hspace{-.1cm}\omega,\overline{z}^{p}_{k}) \\ \max(A^{g\dagger +} \underline{\alpha}_{k}\hspace{-.1cm}-\hspace{-.1cm}A^{g\dagger ++} \overline{\alpha}_{k}\hspace{-.1cm}-\hspace{-.1cm}\omega,\underline{z}^{p}_{k}) \end{bmatrix}\hspace{-.1cm}, \label{eq:global_update}
\end{align}
 This allows us to use the \emph{global} parallel affine abstraction slope $A^g$ for the stability analysis as follows. 
Dropping index $i$ in \eqref{eq:t}--\eqref{eq:alpha} and defining $\Delta^z_k \triangleq \overline{z}_k -\overline{z}_k $ (and similarly for $\Delta^{z^p}_k,\Delta^g_e,\Delta^f_e,\Delta^h_e,\Delta^ {\alpha}_k,\Delta^t_k$), \eqref{eq:iter_update} implies that $\forall D_1 \in \mathbb{D}_{n+p}$ 
\begin{align}
\nonumber \Delta^z_k &\leq \min (|A^{g\dagger}|\Delta^ {\alpha}_k+2 \kappa \mathbf{r},\Delta^{z^p}_k) \\ 
&\leq D_1(|A^{g\dagger}|\Delta^ {\alpha}_k+2 \kappa \mathbf{r})+(I-D_1)\Delta^{z^p}_k, \label{eq:first_ineq}
\end{align} 
   where the second inequality follows from generalization of the fact that $\min(a,b) \leq \lambda a + (1- \lambda)b, \forall a,b \in \mathbb{R},\lambda \in [0,1]$. Moreover, \eqref{eq:t}--\eqref{eq:alpha} and similar reasoning imply that $\forall D_2 \in \mathbb{D}_l$: 
 \begin{align}
\nonumber \Delta^{\alpha}_k &\leq \min (|W^{g}|\Delta v+\Delta^g_e,|A^g| \Delta^{z^p}_k) \\ 
&\leq D_2(|W^{g}|\Delta v+\Delta^g_e)+(I-D_2)|A^g|\Delta^{z^p}_k. \label{eq:sec_ineq}
\end{align}
On the other hand, by similar arguments, it follows from \eqref{eq:x_prop_abst}--\eqref{eq:x_prop_up} that $\forall D_3 \in \mathbb{D}_n $, 
 \begin{align}
\label{eq:third_ineq} \hspace{-.2cm}\Delta^{z^p}_k &\hspace{-.15cm}\leq \hspace{-.15cm} \begin{bmatrix} D_3(|A^{f}|\Delta^z_{k-1} \hspace{-.1cm} + \hspace{-.1cm} |W^f| \Delta w \hspace{-.1cm}+\hspace{-.1cm}\Delta^f_e)\hspace{-.1cm}+\hspace{-.1cm}(I\hspace{-.1cm}-\hspace{-.1cm}D_3) \Delta^f_{k-1} \\ |A^h| \Delta^z_{k-1}\hspace{-.1cm}+\hspace{-.1cm}|W^h| \Delta w\hspace{-.1cm}+\hspace{-.1cm}\Delta^h_ e \hspace{-.15cm}\end{bmatrix}\hspace{-.1cm}, 
\end{align} 
where $\Delta^f_{k-1} \triangleq f_d(\overline{\zeta}_{k-1},\underline{\zeta}_{k-1})-f_d(\underline{\zeta}_{k-1},\overline{\zeta}_{k-1})$. Furthermore, by Lemma \ref{lem:lip-dec}, $\Delta^f_{k-1} \leq (|A^f|+2C^f_z)\Delta^z_{k-1}+(|W^f|+2C^f_w)\Delta w+\Delta^f_ e$, with $C^f =\begin{bmatrix} C^f_z & C^f_u & C^f_w \end{bmatrix}$ given in \eqref{eq:decompconstruct}.
This, in addition to \eqref{eq:first_ineq}--\eqref{eq:third_ineq}, Proposition \ref{prop:bounding} and non-negativity of both sides of all the inequalities, imply that
\begin{align} 
\Delta^z_k &\leq \mathcal{A}^g(D_1,D_2)\mathcal{A}^{f,h}(D_3)\Delta^z_{k-1} \label{eq:error_dynamics}\\
\nonumber &+\Delta^g(D_1,D_2)+\mathcal{A}^g(D_1,D_2) \Delta^{f,h}(D_3)+2 \kappa D_1\mathbf{r},
\end{align}
for $(D_1,D_2,D_3) \in \mathbb{D}^* \triangleq \mathbb{D}_{n+p} \times \mathbb{D}_l \times \mathbb{D}_n $, where $\mathcal{A}^g(D_1,D_2) \triangleq D_1|A^{g\dagger}|D_2 |A^g|+(I-D_1)$, $\mathcal{A}^{f,h}(D_3) \triangleq \begin{bmatrix} (|A^f|+2(I-D_3)C^f_z)^\top & |A^h|^\top \end{bmatrix}^\top$, $\Delta^g(D_1,D_2) \triangleq D_1|A^{g\dagger}|D_2(|W^g|\Delta v+\Delta^g_e)$ and $\Delta^{f,h}(D_3) \triangleq \begin{bmatrix} ((|W^f| \hspace{-.1cm}+ \hspace{-.1cm} 2(I-D_3)C^f_w) \Delta w \hspace{-.1cm}+ \hspace{-.1cm} \Delta^f_e)^\top & (|W^h|\Delta w \hspace{-.1cm} + \hspace{-.1cm} \Delta^h_e )^\top \end{bmatrix}^\top $. Since $\kappa$ can be infinitely large, in order to to make the right hand side of \eqref{eq:error_dynamics} finite n finite time, we choose $D_1 \in \mathbb{D}_{n+p}$ such that $D_1\mathbf{r}=0$, i.e., $D_{1,i,i}=0 \ \text{if} \ r(i)=1, \forall i \in \{1\dots n+p\}$. Then, by the \emph{Comparison Lemma} \cite{khalil2002nonlinear}, it suffices for uniform boundedness of $\{\Delta^z_k\}_{k=0}^{\infty}$ that the following dynamic system be stable:
\begin{align} \label{eq:eq:error_dynamics2}
\Delta^z_k &= \mathcal{A}^g(D_1,D_2)\mathcal{A}^{f,h}(D_3)\Delta^z_{k-1}+\tilde{\Delta}(D_1,D_2), 
\end{align}
 where the error term $\tilde{\Delta}(D_1,D_2) \triangleq \Delta^g(D_1,D_2)+\mathcal{A}^g(D_1,D_2) \Delta^{f,h}(D_3)$ is a bounded disturbance. This implies that the system \eqref{eq:eq:error_dynamics2} is stable (in the sense of uniform stability of the interval sequnces) if and only if the matrix $\mathcal{A}(D_1,D_2,D_3) \triangleq \mathcal{A}^g(D_1,D_2)\mathcal{A}^{f,h}(D_3)$ is (non-strictly) stable for at least one choice of $(D_1,D_2,D_3) $, equivalently \eqref{eq:stability}--\eqref{eq:stab_constr} should hold. \qed
\subsection{Proof of Lemma \ref{lem:convergence}}
The proof is straight forward by Proposition \ref{prop:bounding}, applying \eqref{eq:error_dynamics} iteratively, the fact that by Theorem \ref{thm:stability}, $\mathcal{A}(D_1,D_2,D_3)$ is a stable matrix for any tuple of $(D_1,D_2,D_3)$ that is a solution of \eqref{eq:stability}--\eqref{eq:stab_constr} and triangle inequality. \qed
\end{document}